\documentclass[%
 aip,
 amsmath,amssymb,
reprint,%
]{revtex4-2}

\usepackage{graphicx}
\usepackage{dcolumn}
\usepackage{bm}

\usepackage[utf8]{inputenc}
\usepackage[T1]{fontenc}
\usepackage{mathptmx}
\usepackage{etoolbox}
\usepackage{color}
\usepackage{ulem}

\makeatletter
\def\@email#1#2{%
 \endgroup
 \patchcmd{\titleblock@produce}
  {\frontmatter@RRAPformat}
  {\frontmatter@RRAPformat{\produce@RRAP{*#1\href{mailto:#2}{#2}}}\frontmatter@RRAPformat}
  {}{}
}%
\makeatother
\begin{document}

\preprint{AIP/123-QED}

\title{A characteristics-based method for shock-ramp data analysis}

\author{Jingxiang Shen}
 \email{shenjingxiang93@pku.edu.cn}
 \affiliation{Center for Applied Physics and Technology, and College of Engineering, Peking University, Beijing 100871, China}
\author{Wei Kang}
 \affiliation{Center for Applied Physics and Technology, and College of Engineering, Peking University, Beijing 100871, China}
\date{\today}

\begin{abstract}
For the data analysis problem of shock-ramp compression, i.e., ramp compression after a relatively strong initial shock, a characteristics-based method that strictly deals with the initial hydrodynamic shock is described in detail. Validation of this analysis method using simulated shock-ramp data generated by molecular dynamics and one-dimensional radiation hydrodynamic code is also presented.
\end{abstract}

\maketitle

\section{Introduction}
Shock-ramp loading, which means applying an initial shock prior to ramp loading, is a widely used dynamic compression method in the study of high-pressure physics. Form a scientific point of view, it provides an alternative way of probing off-Hugoniot high-pressure states of matter besides the "pure" ramp compression. And in technical aspects, it allows bypassing the extremely complicated "lower-pressure" region dominated by elastic-plastic response and structural phase transitions, thus making the experiment results easier to interpret.

For shock-free "pure" ramp compression, the experimental and data analysis techniques have been well developed for decades \cite{Aidun_1991, Maw2006, Rothman_2005}. It relies on the measurements of free surface or window interface velocity profiles, assisted by some back-calculation methods that reversely construct the isentropic flow field.

Such a back-calculation step is also needed in shock-ramp experiments. But the relatively strong initial shock makes the entire flow field no longer isentropic. Those back-calculation methods initially developed for shock-free ramp experiments would be erroneous when being applied to shock-ramp data. A special treatment on the initial shock is thus needed.
 
There have been many published results on laser-driven shock-ramp experiments over the years\cite{Smith_2018,Fratanduono_2020}. But most of the publications are primarily focus on scientific issues rather than the details of data analysis methods. For the back-calculation methods used therein\cite{Rothman_testproblem2021}, the citations are often traced back to the earliest literature of "pure" ramp compression\cite{Hayes2001_Report,Hayes2002,Maw2006,Rothman_2005} where the special treatment on the initial shock has not been included.

In this manuscript, a method for treating the initial shock rigorously in the context of characteristics-based back-calculation method is provided in detail. It is then verified on the simulated shock-ramp data generated by non-equilibrium molecular dynamics (MD) and one-dimensional radiation hydrodynamic code. The method proposed here is ready to use.

\section{The basic shock-free case}
\label{section2}

\begin{figure}[thb]
\includegraphics{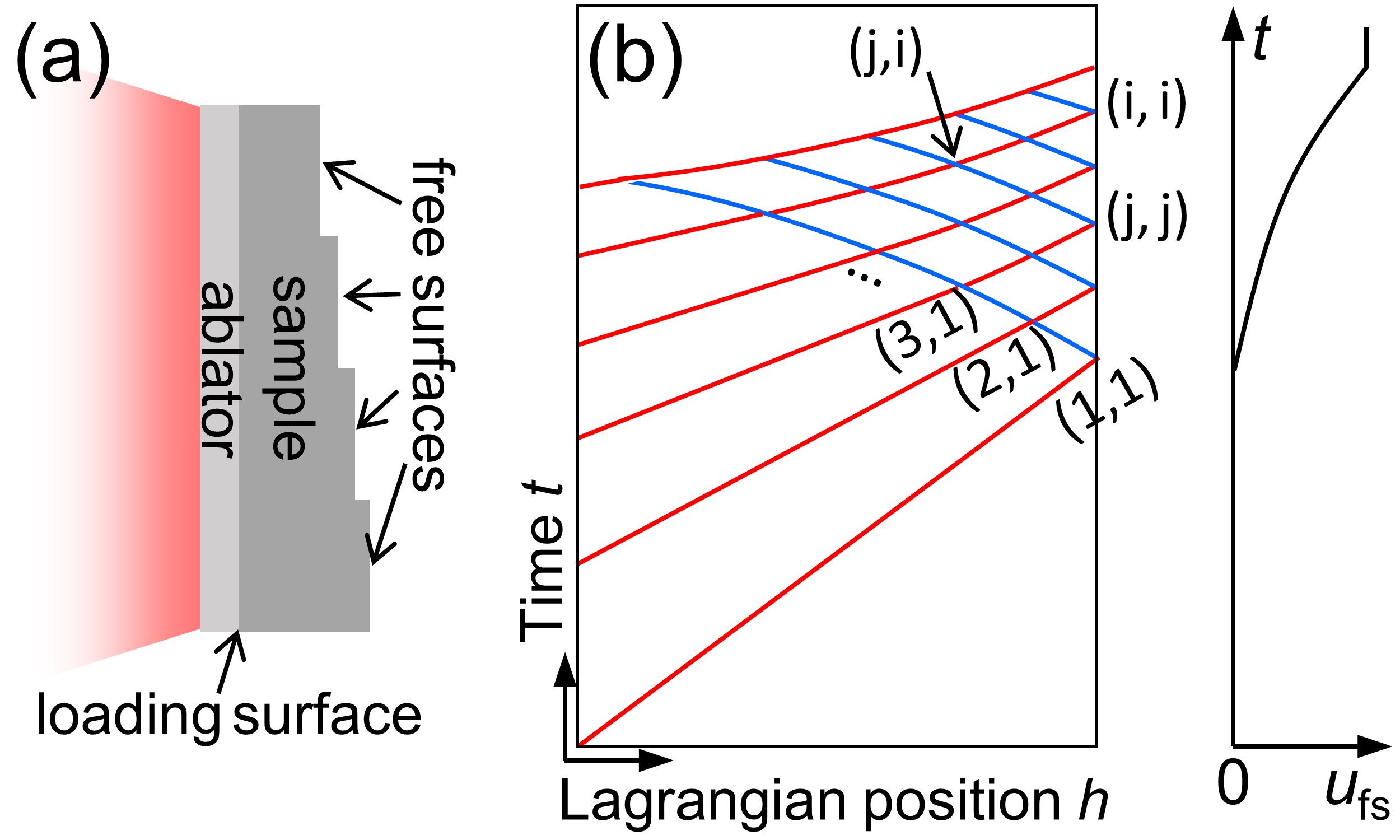}
\caption{\label{fig0} (a) An illustration of the typical setup of laser driven ramp/shock-ramp experiments relevant to this manuscript. (b) An illustration of the characteristics. The intersect of the i-th right-going (compression) wave and the j-th left-going (rarefaction) wave is labeled as (i,j).}
\end{figure}

Let's start by repeating the iterative characteristic approach in the basic shock-free scenarios. This method is also known as the iterative Lagrangian analysis\cite{Rothman_2005} and has been widely used over the years.

The isentropic hydrodynamic equations, which means that a common $P(\rho)$ relation is shared across the entire flow field thus making the energy equation to decouple, can be written in the Lagrangian coordinates as follows.
\begin{equation}
\label{eq:hydro_eqn}
\frac{\partial v}{\partial t}=\frac{\partial u}{\partial h};\ \ \frac{\partial u}{\partial t}=c_L^2\frac{\partial v}{\partial h}
\end{equation}
Here, $h$ stands for the lagrangian position, $u$ is the fluid velocity, $v=\rho_0/\rho$ is the inverse of compression ratio, and $c_L=c\rho/\rho_0$ is the Lagrangian sound velocity. Multiply both ends of the mass equation by $c_L$, then add or subtract the momentum equation to obtain
\begin{equation}
\left(\frac{\partial}{\partial t} \pm c_L\frac{\partial}{\partial h}\right)u \mp \left(\frac{\partial}{\partial t} \pm c_L\frac{\partial}{\partial h}\right)v = 0
\end{equation}
Defining the full derivatives along the right ("+") or left ("-") going characteristics 
\begin{equation}
\left(\frac{D}{Dt}\right) = \ \frac{\partial}{\partial t} \pm c_L \frac{\partial}{\partial h}
\end{equation}
then, eqn.~\ref{eq:hydro_eqn} turns to
\begin{equation}
\left(\frac{D}{Dt}\right)u = \pm c_L \left(\frac{D}{Dt}\right)v
\end{equation}
Performing integral along the right going characteristics gives
\begin{equation}
\label{eq:intR}
u_2-u_1 = \int_{\rho_1}^{\rho_2}c_L dv = -(a_2-a_1)
\end{equation}
and along the left going characteristics,
\begin{equation}
\label{eq:intL}
u_2-u_1 = -\int_{\rho_1}^{\rho_2}c_L dv = a_2-a_1
\end{equation}
Here, we have defined a state quantity with the dimension of velocity along the isentrope:
\begin{equation}
a(\rho) = \rho_0\int_{\rho}^{\rho_0}c_L/\rho^2 d\rho
\end{equation}
which is essentially the equivalent particle velocity achieved by applying a series of simple compression wave on a initially stationary, zero-pressure sample with initial density $\rho_0$.

Characteristics of a typical shock-free ramp loading case with free right boundary is sketched as Fig.~\ref{fig0}~(b). The right going characteristics are colored red and the left going ones are colored blue. Applying eqns.~\ref{eq:intR} and \ref{eq:intL} for the characteristics connecting points $(j,j)$, $(j,i)$ and $(i,i)$, we have
\begin{equation}
u_{ji}-u_{jj}=a_{ji}-a_{jj}; \ \ \ 
u_{ii}-u_{ji}=-a_{ii}+a_{ji}
\end{equation}
Since the right boundary is the free surface (window layer is not considered in this manuscript), we have $a_{ii}=a_{jj}=a(\rho_0)=0$. Therefore,
\begin{equation}
\label{eq:uijaij}
u_{ji}=\frac{1}{2}\left(u_{ii}+u_{jj}\right); \ \ \ 
a_{ji}=\frac{1}{2}\left(u_{ii}-u_{jj}\right)
\end{equation}
The $u$ and $a$ values on each $(j,i)$ point can thus be determined immediately using the measured free surface velocity profile.

The iteration steps go as follows. Starting from an initial guess of $P(\rho)$, the corresponding $c_L(a)$ relation would be
\begin{equation}
c_L(\rho)=\frac{\rho c}{\rho_0}; \ \ \ 
a(\rho)=\int_{\rho_0}^{\rho}\frac{c}{\rho}d\rho
\end{equation}
Substituting the $a_{ij}$ value given by eqn.~\ref{eq:uijaij} into the current $c_L(a)$ relation, we have the $c_L$ at each $(j,i)$ point, which are the local slopes of the $+/-$ characteristic lines. The space-time coordinates of the entire characteristics network can thus be solved for each planar steps with different thickness.
Then we can find the "corrected" free surface arrival times of the simple compression waves as if they were not distorted by the reflected rarefaction waves, from which their wave velocities $c_L$ can be fitted by these wave transit times. This fitted results provide an updated $c_L(a)$ relation, and these processes can be iterated till convergence.

The longitudinal stress $P$ and density $\rho$ can be found straightforwardly from the converged $c_L(a)$ relation. Starting from the definition of sound velocity $c^2=dP/d\rho$, we have
\begin{equation}
\label{eq:Prho01}
P=\int_{\rho_0}^{\rho}c^2 d\rho=\int_{\rho_0}^{\rho}\rho c d\rho
=\rho_0\int_{0}^{a}c_L da
\end{equation}
Here we have used the mass conservation relation of simple compression waves $c d\rho=\rho da$. On the other hand, $c d\rho=\rho da$ can be rewritten as $-\rho_0 d(1/\rho)=da/c_L$; Performing integration of both ends gives the expression of density $\rho$
\begin{equation}
\label{eq:Prho02}
\rho_0 \left(\frac{1}{\rho_0}-\frac{1}{\rho}\right)=\int_0^{a}\frac{da}{c_L(a)}
\end{equation}

\section{Dealing with the initial shock}
\label{section3}
The difficulty of performing back-calculation on the shock-ramp flow field originates from that the information of $c_L$ below the initial shock Hugoniot point is lost -- all $c_L$ below are overtaken by a single value which is the shock velocity. While on the other hand, the free surface velocities are actually affected by the $c_L(a)$ relation in this region.

\begin{figure}[b]
\includegraphics{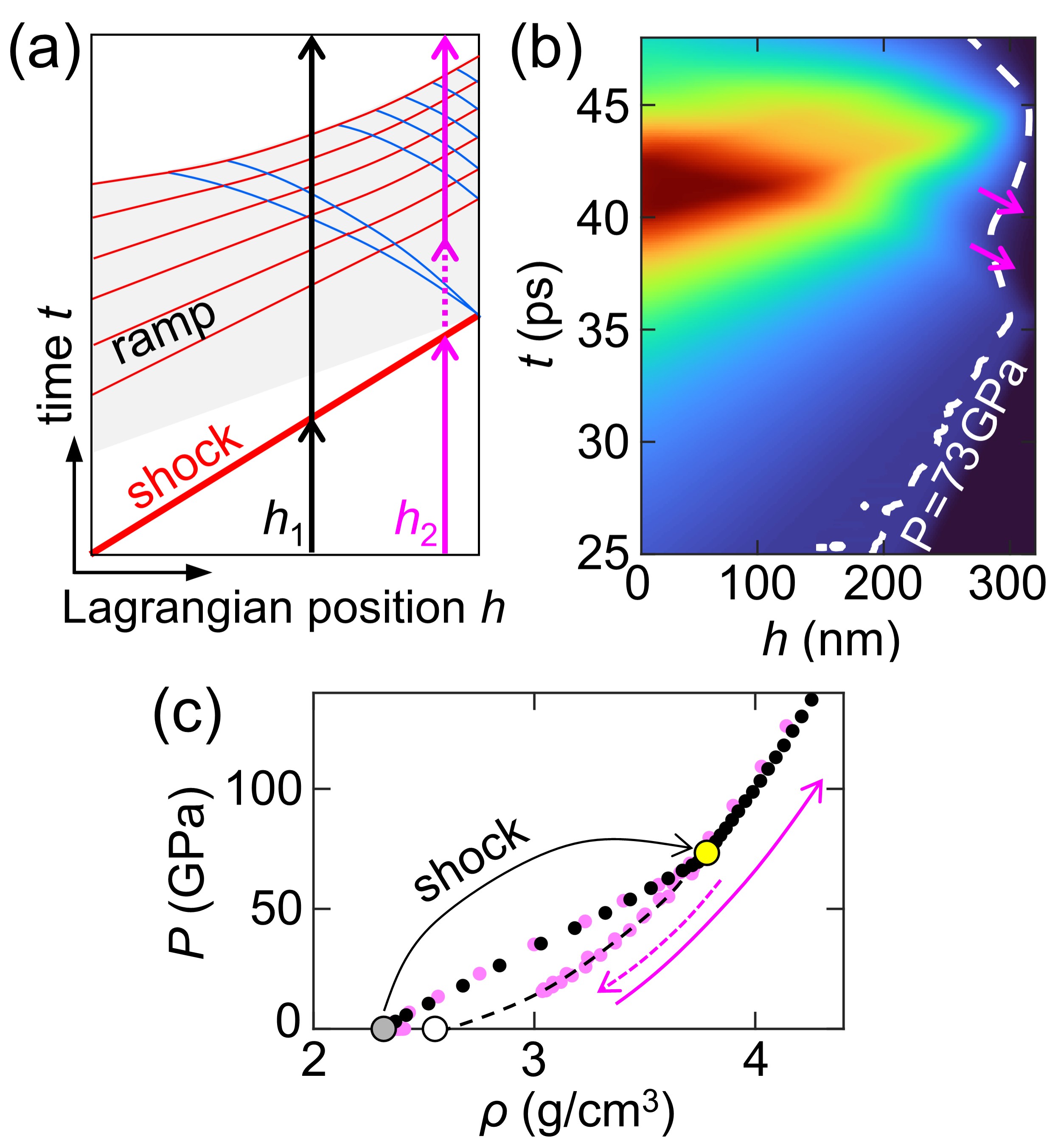}
\caption{\label{fig1} Typical flow field of planar shock-ramp compression. (a) A sketch of the Lagrangian flow field. The flow field is believed to be isentropic if the initial shock is excluded, i.e., in the shaded area hear. (b) The heatmap of longitudinal stress in an MD simulation of shock-ramped silocon. (c) Near the free surface, the sample undergoes release and recompression after the initial shock, during which the stress and density values follow a different "release curve".}
\end{figure}

Fig.~\ref{fig1}~(b) shows the longitudinal stress in a MD simulation case of shock-ramped silicon. A significant portion of the flow field near the free surface, marked by the arrowheads, lies below initial shock Hugoniot pressure, which is about 73.5~GPa in this case. Fig.~\ref{fig1}~(a) provides a schematic illustration. Near the free surface, the sample material is shocked, released, and then recompressed as shown by the magenta line. This feature is also reflected in the pressure-density plot as Fig.~\ref{fig1}~(c). The release and recompression phases follow a different $P(\rho)$ curve than the initial shock compression phase. Note that this low pressure release curve can be nontrivial. For example in the case of shocked silicon here, because atoms are more closely packed in the shock melted (liquid) phase than the initial diamond structure, the released density $\rho_R$ (white dot) is significantly higher than the initial density $\rho_0$ (gray dot). And in other cases lacking such a phase transition, $\rho_R$ can be much lower than $\rho_0$ due to thermal expansion.

MD simulation results like this one indicate that a well-defined $P(\rho)$ relation still exists during the release and recompression phases. This is consistent with the knowledge from hydrodynamics that after being shocked to a fluid-like state, the $P(\rho)$ for subsequent isentropic release and isentropic ramp compression both lie on the isentrope. Therefore, back-calculation based on the isentropic hydrodynamic equations is still possible if the initial shock is excluded from the calculation domain.

Inverse calculation of the flow field relies on a correct description of the release curve below the Hugoniot point. Of course, everything would be solved if a database of equation of states below say 100~GPa were available.
But being realistic, we would like to solve the back-calculation problem in a self-standing manner -- only the free surface velocity profiles and the initial shock Hugoniot state are known. Therefore, we must make reasonable assumptions on the release curve and fit the unknown parameters using the free surface velocity data only.

We assume that the $c_L(a)$ relation between the released state of the free surface (density $\rho_R$, zero pressure) and the Hugoniot state "H" can be approximated by the linear relation:
\begin{equation}
\label{eq:linearcLa}
c_L=c_{LR}(1+\beta a)
\end{equation}
Here, $c_L$ is still defined referring to the pre-shock density $\rho_0 c_L=\rho c$, while the quantity $a$ is defined starting from the released state "R" as $a=\int_{\rho_R}^{\rho} (c/\rho) d\rho$.

Note that the $a$ value of the Hugoniot state $a_H=\int_{\rho_R}^{\rho_H} (c/\rho) d\rho$ is the velocity increment of isentropically unloading to $\rho_R$ starting from $\rho_H$ -- the free surface velocity after shock breakout should be exactly $a_H$ plus the Hugoniot particle velocity $U_H$. So, $a_H$ can be readout from the free surface velocity data by: 1) measure the shock velocity from the shock transit times, 2) find the corresponding $U_H$ by looking up the given $U_s$-$U_p$ Hugoniot of the initial shock, 3) $a_H=u_{fs}-U_H$.

Integrating this hypothetical linear $c_L(a)$ in between the release state "R" and the Hugoniot state "H" gives two constraints related to pressure and density.
\begin{equation}
P_H=\int_0^{a_H} d\rho=\rho_0 c_{LR} \left(a_H+\frac{\beta}{2}a_H^2\right)
\end{equation}
\begin{equation}
\frac{1}{\rho_R}-\frac{1}{\rho_H}=\int_0^{a_H}\frac{da}{\rho_0 c_{LR}}=\frac{\mathrm{ln}(1+\beta a_H)}{\beta \rho_0 c_{LR}}
\end{equation}

The two equations contains a total of 3 unknowns, $c_{LR}$, $\beta$, and $\rho_R$. In principle, the third constraints exists since $c_L$ should be continuous at the Hugoniot point. But for simplicity, we would like to decouple the determination of $c_{LR}$ and $\beta$ from the iterative back-calculation steps. So here we proceed with a hypothetical $\rho_R$ value.
(When doing data analysis with our codes, it is actually very easy to find the proper $\rho_R$ that satisfies the continuity condition after a couple of binary-search trials).
Given $\rho_R$, the unknowns $c_{LR}$ and $\beta$ can be solved immediately. For example we can first eliminate $c_{LR}$:
\begin{equation}
\frac{1}{\rho_R}-\frac{1}{\rho_H}=\frac{\mathrm{ln}(1+\beta a_H)}{\beta} \frac{1+\beta a_H/2}{P_H} a_H
\end{equation}
and then separate the terms containing $\beta$:
\begin{equation}
\frac{P_H}{a_H^2}\left(\frac{1}{\rho_R}-\frac{1}{\rho_H}\right) = \left(\frac{1}{y}+\frac{1}{2}\right) \mathrm{ln}(1+y)
\end{equation}
here we have defined $y:=\beta a_H$. This transcendental equation can be solved easily by numerical interpolation, yielding $\beta$ hence $c_{LR}$. In the subsequent iterative back-calculation of the characteristics, we need to replace at each iteration the $a<a_H$ part of the $c_L(a)$ relation with this linear relation.

After determining the $c_L(a)$ relation below the initial shock Hugoniot point, the remaining problem for back-calculating the characteristics is to deal with the discontinuity in the free surface velocity data. If we try to construct the characteristics backwardly starting from a step-like $u_{fs}$ profile, the result would be something like Fig.~\ref{fig3}~(b) -- accumulative central compression waves converge exactly at the discontinuity point, generating a fan of central rarefaction waves. This is the only way for isentropic flow field to generate discontinuity on the free surface velocity profile.

\begin{figure}[hbt]
\includegraphics{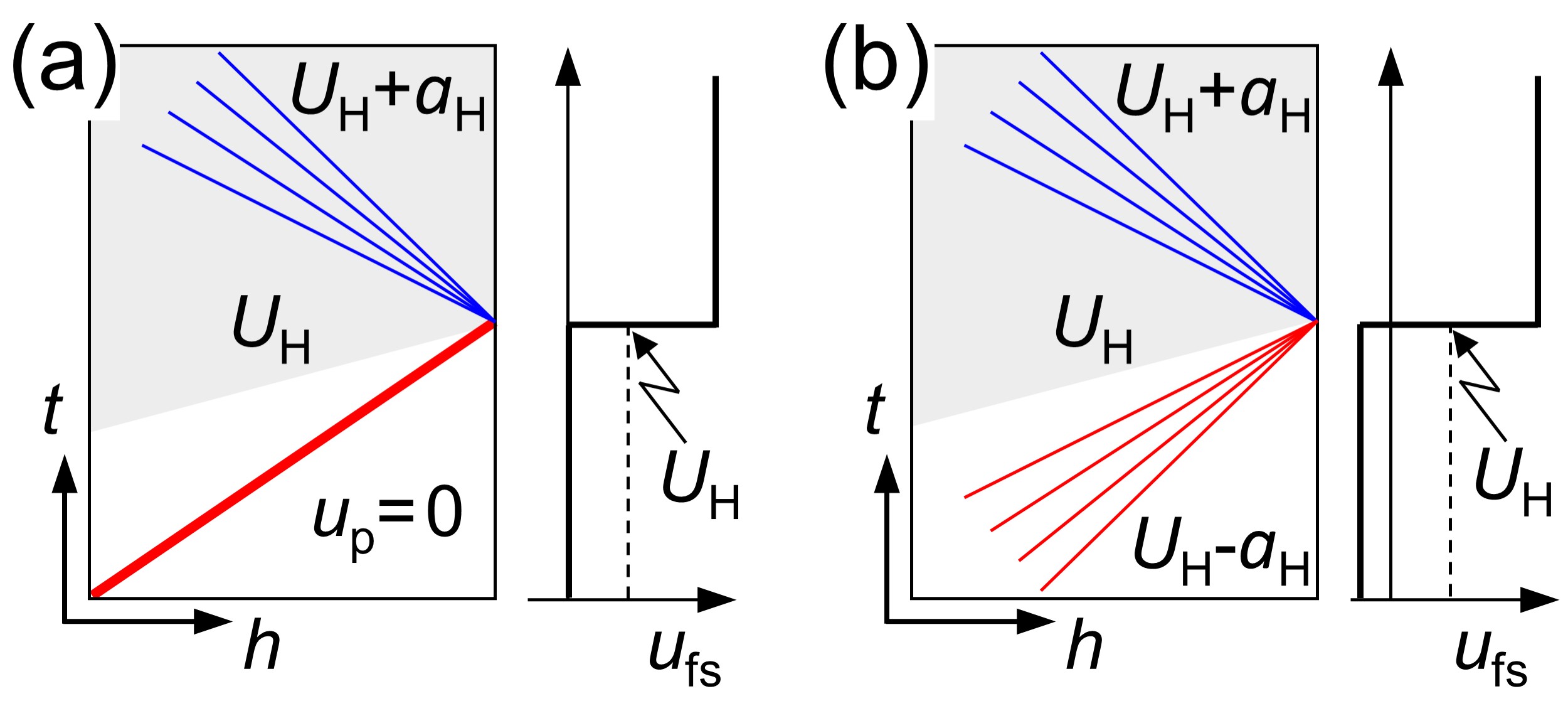}
\caption{\label{fig2} The actual flow field (a) and its isentropic substitution (b) near the shock breakout event. By setting the pre-shock free surface velocity to $U_H-a_H$, the backward characteristics method would give the correct release fan in the part of flow field of interest here (shaded area). In practice since the jump in $u_{fs}$ is not infinitely sharp, a linear stretch is applied to the $u_{fs}<U_H$ part.}
\end{figure}

Of course, this kind of flow field is far from reality. However, by some modifications we can make the central rarefaction waves calculated in this way to effectively simulate the rarefaction waves in reality. The trick is to manually set the pre-shock free surface velocity to $U_H-a_H$, as shown in Fig.~\ref{fig2}~(b). This substitution is fully legitimated since the only differences in flow field appears outside the shaded area. So, the entire network of right and left going characteristics relevant to the simple compression waves are not affected. After this modification, eqn.~\ref{eq:uijaij} would yield the correct values of $a$ and the particle velocity $u_p$ at all $(j,i)$ points.

After solving the space-time coordinates of the characteristics for each of the different sample thicknesses, we can fit the wave velocity $c_L$ of each simple compression wave corresponding to the particle velocity $u_p$. One last thing to be noted is that there exists a shift between the actual particle velocity $u_p$ and the quantity $a$, which is the hypothetical particle velocity if the sample is ramped from the initial density $\rho_R$.
\begin{equation}
a = u_p - U_H + a_H
\end{equation}

So far, the iterative characteristics-based algorithm we developed to solve the $c_L(a)$ relation with the presence of initial shock has been fully described.

For the case of shock-ramp compressed silicon mentioned above, our algorithm has greatly reduced the error of the back-calculated flow field (Fig.~\ref{fig3}) compared with the basic shock-free algorithm. The only remaining error in Fig.~\ref{fig3}~(b) appears near the shock breakout point because our modification on $u_{fs}$. On the one hand this modification corrects the flow field on the large scale, while on the other hand, it introduces some error in the very vicinity as particle velocity on the free surface is manually changed after all.

\begin{figure}[thb]
\includegraphics{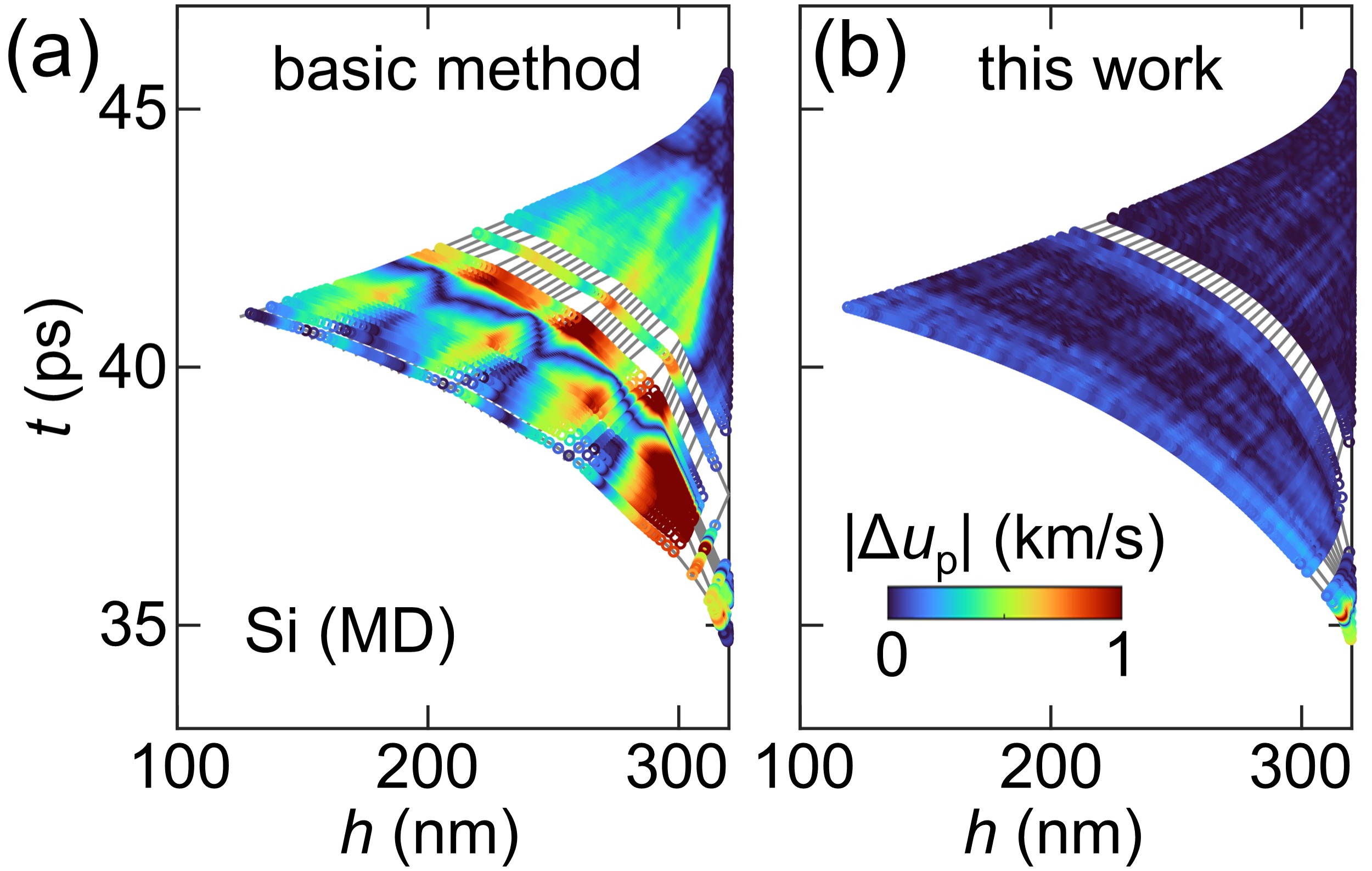}
\caption{\label{fig3} The remaining error after the $c_L(a)$ iteration converges for the case of shock-ramped silicon shown in Fig.~\ref{fig1}~(b). (a) The basic shock-free algorithm. (b) Our algorithm with explicit treatment of the initial shock. The absolute differences in particle velocity are shown as colors. The two panels shares the same color range.}
\end{figure}

The final step is to integrate the converged $c_L(a)$ relation to find pressure and density. With full knowledge of the initial shock, the integration starts from its Hugoniot state.
\begin{equation}
\label{eq:PrhoH1}
P=P_H+\rho_0 \int_{a_H}^{a} c_L da
\end{equation}
\begin{equation}
\label{eq:PrhoH2}
\rho=\left(\frac{1}{\rho_H}-\frac{1}{\rho_0}\int_{a_H}^{a} \frac{da}{c_L}\right)^{-1}
\end{equation}

\section{Validation on simulated data}

\begin{figure*}[p]
\includegraphics{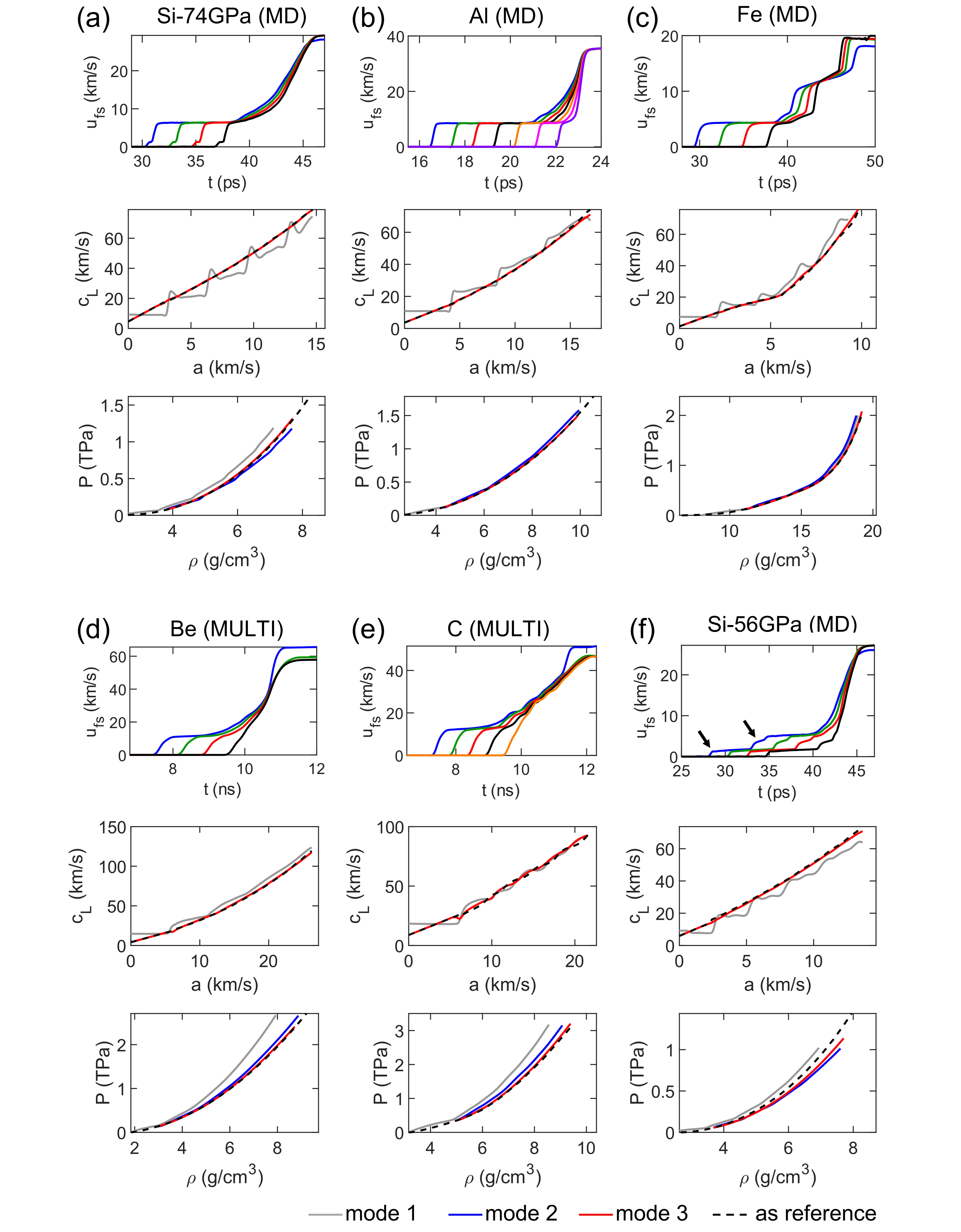}
\caption{\label{fig4} Validation of the proposed back-calculation algorithm in simulation data. Six shock-ramp simulation cases generated by non-equilibrium molecular dynamics or the radiation hydrodynamics code MULTI\cite{RAMIS1988} are presented. For each case, the free surface velocity profiles, the converged $c_L(a)$ relation, and the reconstructed pressure-density curve are shown. Three different analysis modes are compared against each other: mode~1 is the basic iterative characteristics-based algorithm introduced in Section~\ref{section2}; mode~2 is a minimum modification of mode~1 that the final integration step follows eqns.~\ref{eq:PrhoH1}~and~\ref{eq:PrhoH2} rather than eqns.~\ref{eq:Prho01}~and~\ref{eq:Prho02}; mode~3 is the method proposed by this work. Mode~3 works well for panels (a)-(e) where the initial shock is a simple hydrodynamic shock, but not for panel~(f) where the initial shock shows distinctive elastic-plastic two-wave structure.}
\end{figure*}

A total of six simulation cases are presented here. Four of them are generated by MD simulation including the silicon case with 73.5~GPa initial shock shown above in Figs.~\ref{fig1}~and~\ref{fig3}. The shock and ramp waves are generated by the moving left boundary. The other two cases are generated by the one-dimensional radiation hydrodynamics code MULTI\cite{RAMIS1988}, using carefully designed radiation temperature profiles. 

Three different modes of data analysis are discussed. Their results are compared against the "ground truth" $P(\rho)$ curve extracted directly from the simulated flow field. Mode~1 is the basic shock-free mode introduced in Section~\ref{section2}. Mode~2 is an \textit{ad hoc} modification of mode~1, that the integration of pressure and density starts from the Hugoniot point instead of from $\rho_0$, while the iteration for $c_L(a)$ still follows the basic shock-free algorithm. Mode~3 is the method proposed by this work, using the iteration algorithm described in Section~\ref{section3} and eqns.~\ref{eq:PrhoH1}~and~\ref{eq:PrhoH2} for the final integration.

The results are summarized in Fig.~\ref{fig4}. For each of the validation cases, the upper panel shows the free surface velocity profiles generated by simulation. The middle panel is the converged $c_L(a)$ relation in comparison with that derived from the directly measured $P(\rho)$ curve. Note that modes~1~and~2 are both based on the basic shock-free back-calculation algorithm so only one $c_L(a)$ curve is shown. The final pressure-density results for the three analysis modes are given in the lower panel. Also shown as the dashed black line is the "ground truth" $P(\rho)$ extracted directly from the MD or hydrodynamic simulations, serving as the reference here.

To quantify the back-calculation errors of these three analysis modes, we use the averaged deviation in pressure across the density range $\rho_H<\rho<\rho_{max}$ normalized by the averaged pressure:
\begin{equation}
\mathrm{err}=\frac{\left< |P-P_{ref}| \right>_{\rho}}{\left< (P+P_{ref})/2 \right>_{\rho}}
\end{equation}
Their exact values for all three analysis modes on all six validation cases are listed in Table~\ref{tab:1}.

\begin{table}[hb]
    \centering
    \caption{Settings and results of the validation cases}
    \begin{tabular}{l|c|c|c|c|c|c}
         \toprule
         Cases & (a) & (b) & (c) & (d) & (e) & (f) \\
         \hline
         Material & Si & Al & Fe & Be & C & Si \\
         \hline
         Sim. Method & MD & MD & MD & MULTI & MULTI & MD \\
         \toprule
         $\rho_0~(g/cm^3)$ & 2.318 & 2.698 & 8.014 & 1.84 & 3.16 & 2.318 \\
         \hline
         $P_H$~(GPa) & 73.4 & 116.9 & 117.5 & 128 & 334 & 56.5 \\
         \hline
         $U_H~(km/s)$ & 3.5 & 4.0 & 2.0 & 5.06 & 6.17 & 3.0 \\
         \hline
         $a_H~(km/s)$ & 2.86 & 4.46 & 2.33 & 6.112 & 6.105 & 2.40 \\
         \hline
         $\rho_H~(g/cm^3)$ & 3.871 & 4.278 & 10.98 & 2.91 & 4.94 & 3.681 \\
         \hline
         $\rho_R~(g/cm^3)$ & 2.55 & 2.305 & 6.52 & 1.45 & 3.10 & 2.63 \\
         \toprule
         err mode~1~(\%) & 22.96 & 7.40 & 6.53 & 28.92 & 30.05 & 16.37 \\
         \hline
         err mode~2~(\%) & 5.81 & 6.75 & 8.66 & 6.70 & 11.39 & 15.67 \\
         \hline
         err mode~3~(\%) & 1.64 & 0.55 & 0.97 & 1.08 & 2.42 & 11.23 \\
         \toprule
    \end{tabular}
    \label{tab:1}
\end{table}

It's clear that mode~3, i.e., the analysis method proposed by this work, shows vanishing error (on the order of 1\% in pressure) for most of the cases, except for Fig.~\ref{fig4}~(f). This last example is deliberately included to show the range of validity of the hydrodynamic approaches taken by the back-calculation algorithms. In this case, the shock is not strong enough to overtake the elastic precursor and there are actually two successive shocks -- an elastic one and an plastic one. The splitting of the shock front can be seen in the free surface velocity curves (Fig.~\ref{fig4}~(f), arrowheads in the upper panel). Our treatments of the initial shock described in Section~\ref{section3} is developed for a single initial shock therefore failed in this case. 

As a comparison, although the elastic precursor is still visible in Fig.~\ref{fig4}~(a), the elastic and plastic shock fronts travel as nearly the same speed. This makes them to reach the free surface almost simultaneously, and their corresponding rarefaction waves can be viewed as originating from approximately the same shock breakout point. So our treatments of the shock-produced rarefaction waves are still valid.

\section{Conclusion}
We provide in detail a characteristics-based iterative back-calculation method for analyzing shock-ramp loading experiment data. The method treats the initial shock rigorously to the first order in the sense of the $c_L(u_p)$ relation. 
The method is self-standing, in that it relies only on the shock-ramp free surface velocity profiles $u_{fs}(t)$ and the initial shock Hugoniot state quantities $\rho_H$, $U_H$, and $P_H$. No external datasets on the release curves are needed. 
It works well as long as the initial shock can be treated as a single hydrodynamic discontinuity (i.e., no significant elastic-plastic splitting). The method is implemented as a MATLAB package and is ready to use.

\begin{acknowledgments}
We thank Dr. Xiaoxi Duan for helpful discussions. 
\end{acknowledgments}

\section*{Data and Code Availability Statement}
The data and code that support the findings of this study are available from the corresponding author upon reasonable request.

\def\refname{References}
\bibliography{ramp2023B}

\end{document}